\documentclass{jpsj2}
\def\runtitle{Finger-print of Ultra High Energy Cosmic Ray  by Relic
Neutrino Masses in Dark Halos}
\def\runauthor{Daniele {\sc Fargion}, Marco {\sc Grossi}, P.G. {\sc De Sanctis Lucentini},
C. {\sc Di Troia}}

\title{Clustering, Anisotropy, Spectra of Ultra High Energy Cosmic Ray:\\
              Finger-prints of Relic Neutrinos Masses in Dark Halos}

\author{%
Daniele {\sc Fargion}, M. {\sc Grossi}, P.G. {\sc De Sanctis
Lucentini}, C. {\sc Di Troia} }

\inst{%
Physics Department,Rome University 1, Italy \\
INFN Rome, Italy}

\recdate{August 4, 2001}

 \abst{ The Ultra High Energy Cosmic Ray (UHECR), by $\nu_r$-Z
showering in Hot Dark Halos (HDM), should exhibits an energy
spectra and an anisotropy reflecting (also) the relic neutrino
masses and their hierarchical HDM halo clustering. The lighter
are the relic $\nu$ masses, the higher their corresponding Z
resonance energy peaks and their hadronic UHECR tails, the wider
their dark halos and the smaller their clustering density
contrast (and their interaction probability). A {\em twin light}
neutrino mass splitting may reflect to twin Z resonance and into
a complex UHECR spectra modulation ({\em a twin bump}) at the
edge at highest GZK energy cut-off.  Each possible $\nu$ mass
associates a characteristic dark halo size (galactic, local,
super cluster) and its local anisotropy due to our peculiar
position within that dark matter distribution. The expected Z or
WW,ZZ showering into $p$ $\bar{p}$ but also $n$ $\bar{n}$ should
imprint a peculiar matter-anti matter symmetry in observed UHECR
clustering. A $\nu$ HDM halo around a Mpc will allow to the UHECR
$n$ $\bar{n}$ secondary component at $E_n> 10^{20}$ eV (due to Z
decay) to arise playing  a role comparable with the charged $p$
$\bar{p}$ ones. Their un-deflected $n$ $\bar{n}$ (or decayed $p$
$\bar{p}$)  flight is shorter leading to a prompt and hard UHECR
trace pointing toward the original UHECR source direction. The
direct $p$ $\bar{p}$ pairs are split and spread by random
magnetic fields into a more diluted and smeared UHECR signal
around the original source direction. TeVs signals by synchrotron
radiation must also mark the Z-WW Showering. The UHE $\nu$- Z
showering signatures may be already found in recent (and future)
events in AGASA (and Auger) data. The observed hard doublet and
triplets spectra, their time and space clustering already favour
the rising key role of UHECR $n$ $\bar{n}$ secondaries originated
by $\nu$-Z tail shower}

\kword{%
Cosmic Ray, Neutrinos Mass, Dark Halos}

\def\lesssim{\mathrel{\hbox{\rlap{\hbox{\lower4pt\hbox{$\sim$}}}\hbox{$<$}}}}
\def\gtrsim{\mathrel{\hbox{\rlap{\hbox{\lower4pt\hbox{$\sim$}}}\hbox{$>$}}}}

\twocolumn
\begin{document}
\sloppy \maketitle \baselineskip=1.0pc
\section{Introduction}

 Neutrino with a light mass may play a relevant role in solving the puzzle
 of Hot Dark Matter within a hot-cold dark matter (HCDM) scenario. At the
same time their clustering in Galactic, Local dark halos offer the
possibility to overcome the Greisen Zatsepin Kuzmin cut-off
($\gtrsim 4 \cdot 10^{19}\,eV$) (GZK)  at highest energy cosmic
ray astrophysics.\\
These rare events almost  in isotropic spread are probably
originated by blazars AGN, QSRs  in standard scenario, and they
should not come, if originally of hadronic nature, from large
distances because of the electromagnetic dragging friction of
cosmic 2.75 K BBR and of the lower energy diffused inter-galactic
radio backgrounds. Indeed as noted by Greisen, Zatsepin and
Kuzmin \cite{G},  \cite{ZK}, proton and nucleon mean free path at
E $> 5 \cdot 10^{19} \,EeV$ is less than 30 $Mpc$ and
asymptotically nearly ten $Mpc$.; also gamma rays at those
energies have even shorter interaction length ($10 \,Mpc$) due to
severe opacity by electron pair production via microwave and radio
background interactions \cite{Proth1997}. Nevertheless these
powerful sources (AGN, Quasars, GRBs) suspected to be the unique
source able to eject such UHECRs, are rare  at nearby distances
($\lesssim 10 \div 20 \, Mpc$, as for nearby $M87$ in Virgo
cluster); moreover there are not nearby $AGN$ in the observed
UHECR arrival directions.  Strong and coherent
galactic\cite{Proth1997} or extragalactic\cite{Farrar} magnetic
fields, able to bend such UHECR (proton, nuclei) directions are
not really at hand. The needed coherent lengths and strength are
not easily compatible with known cosmic data on polarized Faraday
rotation. Finally in latter scenario the same contemporaneous
ultra-high energy $ZeV$ neutrons born, by photo-pion production
on BBR, may escape the magnetic fields bending  and should keep
memory of the primordial nearby (let say $M87$) arrival
direction, leading to (unobserved) in-homogeneities toward the
primary source. Finally secondaries EeV photons (by neutral pion
decays) should also abundantly point and cluster toward the same
nearby $AGN$
sources \cite{El},\cite{Sigl}  contrary to (never observed) $AGASA$ data.\\
Another solution of the present GZK puzzle, the Topological
defects ($TD$), assumes as a source, relic heavy particles of
early Universe; they are imagined diffused as a Cold Dark Matter
component, in galactic or Local Group Halos. Nevertheless the
$TD$ fine tuned masses and ad-hoc decays are unable to explain the
growing evidences of doublets and triplets clustering in $AGASA$~
$UHECR$ arrival data. In this scenario there have been recent
suggestions \cite{Blasi} for an unexpected population of such 500
compact dark clouds of $10^8 M_{\odot}$, each one made by such
dark $TD$ clusters, spread in our galactic halo; these dark
clouds are assumed  nevertheless,  not correlated to luminous
known galactic halo, disk, globular clusters and center
components. We found all these speculations not plausible. On the
other side there are possible evidences of self-correlation
between UHECR arrival directions with far Compact
Blazars\cite{AGASA} at cosmic distance well above GZK cut-off \cite{Tinyakov}.\\
Therefore the solution of UHECR puzzle based on primary Extreme
High Energy (EHE) neutrino beams (from AGN) at $E_{\nu} > 10^{21}$
eV and their undisturbed propagation from cosmic distances up to
nearby calorimeter (made by relic light $\nu$ in dark galactic or
local dark halo \cite{FarSal97} \cite{FarSal99} \cite {Weiler}
\cite{Yoshida}) is still, in our opinion, the most favorite
conservative solution for the GZK puzzle. Interestingly new
complex scenarios for each neutrino mass spectra are then opening
and important signature of UHECR Z,WW showering must manifest
in observed anisotropy and space-time clustering.\\

\section{UHE neutrino scattering on $\nu_r$ neutrino masses}

If relic neutrinos have a mass   larger than their thermal energy
(1.9 $K^0$) they may cluster in galactic or Local Group halos; at
eVs masses the clustering seem very plausible and it may play a
role in dark hot cosmology\cite{Fargion83}. Their scattering with
incoming extra-galactic EHE neutrinos determine high energy
particle cascades which could contribute or dominate the observed
UHECR flux at $GZK$ edges. Indeed the possibility that neutrino
share a little mass has been reinforced by Super-Kamiokande
evidence for atmospheric neutrino anomaly via $\nu_{\mu}
\leftrightarrow \nu_{\tau}$ oscillation. An additional evidence
of neutral lepton flavour mixing has been very recently reported
also by Solar neutrino experiment (SNO,Gallex,K2K). Consequently
there are at least two main extreme scenario for hot dark halos:
either $\nu_{\mu}\, , \, \nu_{\tau}$ are both extremely light
($m_{\nu_{\mu}} \sim m_{\nu_{\tau}} \sim \sqrt{(\Delta m)^2} \sim
0.05 \, eV$) and therefore hot dark neutrino halo is very wide
and spread out to local group clustering sizes (increasing the
radius but loosing in the neutrino density clustering contrast),
or $\nu_{\mu}, \nu_{\tau}$ may share degenerated ($eV$ masses)
split by a very tiny different values.  In the latter fine-tuned
neutrino mass case ($m_{\nu}\sim 0.4 eV-1.2 eV$) (see Fig,2 and
Fig.3) the Z peak $\nu \bar{\nu}_r$ interaction \cite{FarSal97,
FarSal99} \cite{Weiler, Yoshida} will be the favorite one; in the
second case (for heavier non constrained neutrino mass ($m_{\nu}
\gtrsim 3 \, eV$)) only a $\nu \bar{\nu}_r \rightarrow
W^+W^-$\cite{FarSal97,FarSal99}, and the additional $\nu
\bar{\nu}_r \rightarrow ZZ$ interactions, (see the cross-section
in Fig.1)\cite{Far2001} considered here will be the only ones
able to solve the GZK puzzle. Indeed the relic neutrino mass
within HDM models in galactic halo near $m_{\nu}\sim 4 eV$,
corresponds to a lower and $Z$ resonant incoming energy

\begin{equation}
 {{E_{\nu} =  {\left(
\frac{4eV} {\sqrt{{{m_{\nu}}^2+{p_{\nu}^2}}}} \right)} \cdot
10^{21} \,eV.} \nonumber}
\end{equation}

   This resonant incoming neutrino energy is unable to overcome GZK energies while it is
     showering mainly a  small energy fraction into nucleons ($p,\bar{p}, n, \bar{n}$),
   (see $Tab.1$ below), at energies $E_{p}$ quite below. (see $Tab.2$
   below).

\begin{equation}
 {{E_{p} =  2.2 {\left(
\frac{4eV} {\sqrt{{{m_{\nu}}^2+{p_{\nu}^2}}}} \right)} \cdot
10^{19} \,eV.} \nonumber}
\end{equation}

   Therefore too heavy ($> 1.5 eV$) neutrino mass are not fit to
   solve GZK by Z-resonance while WW,ZZ showering as well as t-channel showering
   may naturally keep open the solution.
   In particular the overlapping of both the Z and the WW, ZZ
   channels described in fig.1, for $m_{\nu} \simeq 2.3 eV$ while
   solving the UHECR above GZK they must pile up (by Z-resonance
   peak activity) events at $ 5 \cdot 10^{19} eV$, leading to a bump in
   AGASA data. There is indeed a first marginal evidence of such a UHECR bump
    in AGASA and Yakutsk data that may stand for this interpretation.
     More detailed data are  needed to verify such conclusive  possibility.
      Similar result regarding the fine tuned relic mass
    at $0.4 eV$ and $2.3 eV$, (however ignoring the WW ZZ and
    t-channels and invoking very hard UHE neutrino spectra) have been
    independently reported recently \cite{Fodor2001}.\\
   Most of us  consider cosmological light relic neutrinos in Standard Model
   at non relativistic regime   neglecting any relic neutrino momentum ${p_{\nu}} $ term.
     However, at lightest mass values
   the momentum may be comparable to the relic mass; moreover
   the spectra may reflect additional relic neutrino-energy injection which are feeding
   standard cosmic relic neutrino   at energies much above the same neutrino mass.
    Indeed there may  exist, within or beyond Standard Cosmology,
      a relic neutrino component due to stellar,
   Super Nova, GRBs, AGN past activities, presently  red-shifted
    into a  KeV-eV spectra, piling into a relic neutrino grey-body  spectra.
     Therefore  it is worth-full to keep the most general
      mass and momentum term in the target relic neutrino spectra.
      In this windy ultra-relativistic neutrino  cosmology, eventually
      leading to a neutrino radiation dominated Universe, the
      halo size to be considered is nearly coincident with the GZK one defined by
      the energy loss lenght for UHECR nucleons ($\sim 20 Mpcs$).
       Therefore the isotropic UHECR behaviour
      is guaranteed but a puzzle related to uniform source
      distribution  seem to persist. Nevertheless the UHE neutrino-
      relic neutrino scattering \textit{do not} follow a flat
      spectra as shown in figure 2, (as well as any hypothetical $\nu$ grey
      body spectra). This leave  open the opportunity to have a
      relic relativistic neutrino component at eVs energies as
      well as the observed  non uniform UHECR spectra. This case is similar to the
       case of a very light neutrino mass much below $0.1$ eV.\\

As we noticed above, relic neutrino mass above a few eVs in  HDM
halo \textit{are not} consistent with naive Z peak; higher
energies interactions ruled by WW,\cite{Enq,FarSal99} ZZ
cross-sections \cite{Far2001} may nevertheless solve the GZK
cut-off. In this regime there will be also possible to produce by
virtual W exchange, t-channel, $UHE$ lepton pairs, by $\nu_i
\bar{\nu}_j\rightarrow l_i\bar{l}_j$,
leading to additional electro-magnetic showers injection.\\
As we shall see these important and underestimated signal will
produce UHE electrons whose final trace are TeVs synchrotron
photons.
 The hadronic tail of the Z or $W^+ W^-$ cascade maybe
 the source of final  nucleons $p,\bar{p}, n, \bar{n}$ able to explain UHECR events observed by
Fly's Eye and AGASA \cite{AGASA} and other detectors. The same
$\nu \bar{\nu}_r$ interactions are source of Z and W that decay in
rich shower ramification. The average energy deposition for both
gauge bosons among the secondary particles is summarized in Table
1A below.

\begin{figure}[h]
\begin{center}
 \includegraphics[width=0.45\textwidth] {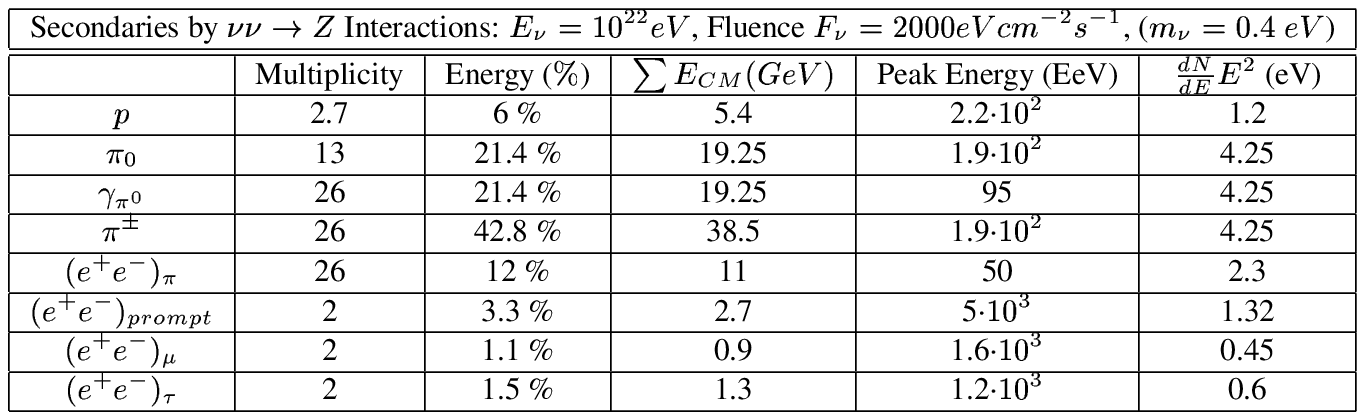}
\end{center}
  \caption{Table 1A: The total detailed energy percentage distribution  into neutrino,
protons, neutral and charged pions and consequent gamma, electron
pair particles both from hadronic and leptonic Z, $WW,ZZ$
channels. We also calculated the elecro-magnetic contribution due
to the t-channel $\nu_i \nu_j$ interactions. We used LEP data for
Z decay and considered W decay roughly in the same way as Z one.
We assumed that an average number of 37 particles is produced
during a Z (W) hadronic decay. The number of prompt pions both
charged (18) and neutral (9), in the hadronic decay is increased
by 8 and 4 respectively due to the decay of $K^0$, $K^{\pm}$,
$\rho$, $\omega$, and $\eta$ particles. (*)We assumed that the
most energetic neutrinos produced in the hadronic decay mainly
come from charged pion decay. SO their number is roughly three
times the number of $\pi$'s. UHE photons are mainly relics of
neutral pions. Most of the $\gamma$ radiation will be degraded
around PeV energies by $\gamma \gamma$ pair production with
cosmic 2.75 K BBR, or with cosmic radio background. The electron
pairs instead, are mainly relics of charged pions and will
rapidly lose energies into synchrotron radiation. The
contribution of leptonic Z (W) decay is also considered and
calculated in the table 1A-1B.} \label{1}
\end{figure}


\section{UHECR  Neutrons  from Z showers}

 Although protons (or anti-protons)
  are the most popular and favorite candidate in
order to explain the highest energy air shower observed, one
doesn't have to neglect the signature of final neutron and
anti-neutrons as well as electrons and photons. Indeed the UHECR
neutrons are produced in Z-WW showering at nearly same rate as
the charged nucleons. Above GZK cut-off energies UHE $n$,$
\bar{n}$, share a life lenght comparable with the Hot Galactic
Dark Neutrino Halo. Therefore they may be an important component
in UHECRs. Moreover prompt UHE electron (positron) interactions
with the galactic or extra-galactic magnetic field or soft
radiative backgrounds may lead to gamma cascades and from PeVs to
TeVs energies.\\
Gamma photons at energies $E_{\gamma} \simeq 10^{20}$ - $10^{19}
\,eV$ may freely propagate through galactic or local halo scales
(hundreds of kpc to few Mpc) and could also contribute to the
extreme edges of cosmic ray spectrum  and clustering
(see also \cite{Yoshida}\cite{Far2001}). \\
The ratio of the final energy flux of nucleons near the Z peak
resonance, $\Phi_p$ over the corresponding electro-magnetic
energy flux $\Phi_{em}$ ratio is, as in tab.1 $e^+ e^-,\gamma$
entrance, nearly $\sim \frac{1}{8}$.  Moreover if one considers
at higher $E_{\nu}$ energies, the opening of WW, ZZ channels and
the six pairs $\nu_e \bar{\nu_{\mu}}$, \, $\nu_{\mu}
\bar{\nu_{\tau}}$, \, $\nu_e \bar{\nu_{\tau}}$ (and their
anti-particle pairs) t-channel interactions leading to  highest
energy leptons, with no nucleonic relics (as $p, \bar{p}$), this
additional injection favors the electro-magnetic flux $\Phi_{em}$
over the corresponding nuclear one $\Phi_p$ by a factor $\sim
1.6$ leading to $\frac{\Phi_p}{\Phi_{em}} \sim \frac{1}{13}$.
This ratio is valid at $WW,ZZ$ masses because the overall cross
section variability is energy dependent. At center of mass
energies above these values, the $\frac{\Phi_p}{\Phi_{em}}$
decreases more because the dominant role of t-channel (Fig1). We
focus here  on Z, and WW,ZZ channels showering in hadrons for GZK
events.
The important role of UHE electron showering into TeV radiation is discussed below.\\
\section{UHE $\nu$ - $\nu_{relic}$ Cross Sections }
Extragalactic neutrino cosmic rays are free to move on cosmic
distances up our galactic halo without constraint on their mean
free path, because the interaction length with cosmic background
neutrinos is greater than the actual Hubble distance. A Hot Dark
Matter galactic or local group halo model with relic light
neutrinos (primarily the heaviest $\nu_{\tau}$ or $ \nu_{\mu} $)
\cite{FarSal99}, acts as a target for the high energy neutrino
beams. The relic number density and the halo size are large
enough to allow the $\nu \nu_{relic}$ interaction. As a
consequence high energy particle showers are produced in the
galactic or local group halo, overcoming the GZK cut-off
\cite{FarSal99}.  There is an upper bound density clustering for
very light Dirac fermions due to the maximal Fermi degenerancy
whose adimensional density contrast is $\delta\rho \propto
m_{\nu}^3$,  while one finds \cite{Fargion83}  that  the neutrino
free-streaming halo grows only as $\propto m_{\nu}^{-1}$.
Therefore the overall interaction probability grows $ \propto
m_{\nu}^{2} $, favoring heavier non relativistic (eVs) neutrino
masses. In this frame above few eV neutrino masses only WW-ZZ
channel are operative. Nevertheless the same lightest relic
neutrinos may share higher Local Group velocities (thousands
$\frac{Km}{s}$) or even nearly relativistic speeds and it may
therefore compensate the common density bound:

\begin{equation}
 n_{\nu_{i}}=1.9\cdot10^{3}
\left( \frac{m_{i}}{0.1eV}\right)  ^{3}\left(
\frac{v_{\nu_{i}}}{2\cdot10^{3}\frac{Km}{s} }\right)  ^{3}
\end{equation}



 From the cross section side there are three main interaction processes that
 have to be considered  leading to nucleons in the
of EHE and relic neutrinos scattering.

 {\bf channel 1.} $\;$ The
$\nu \nu_r\rightarrow Z \rightarrow \, $
 annihilation at the Z resonance.

{\bf channel 2.} $\nu_{\mu} \bar{\nu_{\mu}} \rightarrow W^+ W^-$
or $\nu_{\mu} \bar{\nu_{\mu}} \rightarrow Z Z$ leading to
hadrons, electrons, photons, through W and Z decay.

 {\bf channel 3.} The $\nu_e$ - $\bar{\nu_{\mu}}$, $\nu_e$ -
$\bar{\nu_{\tau}}$, $\nu_{\mu}$ - $\bar{\nu_{\tau}}$ and
antiparticle conjugate interactions of different flavor neutrinos
mediated in the $t$-channel by the W exchange (i.e. $\nu_{\mu}
\bar{\nu_{\tau_r}} \rightarrow \mu^- \tau^+ $). These reactions
are sources of prompt and secondary UHE electrons as well as
photons resulting by hadronic $\tau$ decay.

\subsection{The process $\nu_{\tau} \bar{\nu_{\tau}} \rightarrow Z $}
 The interaction of neutrinos of the
same flavor can occur via a Z exchange in the $s$-channel
($\nu_i\bar{\nu}_{i_r}$ and charge conjugated). The cross section
for hadron production in $\nu_i\bar{\nu}_i\rightarrow
Z^*\rightarrow hadrons$ is
\begin{equation}
\sigma_Z(s)=\frac{8\pi s}{M_Z^2}\frac{\Gamma(Z^o\rightarrow
invis.) \Gamma(Z^o\rightarrow hadr.)}{(s-M_Z^2)^2+M_Z^2
\Gamma_Z^2}
\end{equation}
where $\Gamma(Z^o\rightarrow invis.)\simeq 0.5~GeV$,
$\Gamma(Z^o\rightarrow hadr.)\simeq 1.74~GeV$ and $\Gamma_Z\simeq
2.49~GeV$ are respectively the experimental Z width into invisible
products, the Z width into hadrons and the Z full width
\cite{pdg}. The averaged cross section peak reaches the value ($<
\sigma_Z > = 4.2 \cdot 10^{-32} \, cm^2$).  We assumed here for a
more general case (non relativistic and nearly relativistic relic
neutrinos)  that the averaged cross section has to be extended
over an energy window comparable to half the center of mass
energy. The consequent effective
averaged cross-section is described in Fig.1 as a lower truncated hill curve.\\
A $\nu\nu_r$ interaction mediated in the $s$-channel by the Z
exchange, shows a peculiar peak in the cross section due to the
resonant Z production at $s= M_Z^2$. However, this occurs for a
very narrow and fine-tuned windows of  arrival neutrino energies
${\nu}_{i}$  (and of the corresponding target neutrino masses and
momentum $ \bar{\nu}_{i}$):
\begin{equation}
{E_{\nu_{i}}} =  {\left( \frac{4eV} {\sqrt {{{m_{\nu_
{i}}}^2+{p_{\nu_ {i}}}^2}}} \right)} \cdot 10^{21} \,eV.
\end{equation}

So in this mechanism the energy of the EHE neutrino cosmic ray is
related to the mass of the relic neutrinos, and for an initial
neutrino energy fixed at $E_{\nu} \simeq 10^{22} \, eV$, the Z
resonance requires a mass for the heavier neutral lepton around
$m_{\nu} \simeq 0.4 \, eV$. Apart from this narrow
 resonance peak at $\sqrt{s}= M_Z$, the
asymptotic behaviour of the cross section is proportional to
$1/s$ for $s\gg M_Z^2$.
\\

The $\nu \bar{\nu} \rightarrow Z \rightarrow hadrons$ reactions
have been proposed by \cite {FarSal97} \cite{Weiler}
\cite{Yoshida} with a neutrino clustering on Supercluster,
cluster, Local Group, and galactic halo scale within the few tens
of Mpc limit fixed by the GZK cut-off. Due to the enhanced
annihilation cross-section in the Z pole, the probability of a
neutrino collision is reasonable even for a low neutrino density
contrast $\delta \rho_{\nu} / \rho_{\nu} \geq 10^2$. The potential
wells of such structures might enhance the neutrino local group
density with an efficiency at comparable with observed baryonic
clustering discussed above. In this range the presence of
extended local group halo should be reflected into anisotropy
(higher abundance) toward Andromeda, while a much lighter
neutrino mass may correspond to a huge halo containing even Virgo
and the Super Galactic Plane.

\subsection{The $W^+ W^-$ and $ Z Z $Channels}


The reactions $\nu_{\tau} \bar{\nu_{\tau}} \rightarrow W^+ W^-$,%
 $\nu_{\mu} \bar{\nu_{\mu}} \rightarrow W^+ W^- $,%
 $\nu_{e}  \bar{\nu_{e}} \rightarrow W^+ W^-$, %
 that occurs through the exchange of a Z boson (s channel)\cite{Enq}, has been
previously introduced\cite{FarSal99} in order to explain UHECR as
the Fly's Eye event at 320 Eev detected in 1991 and last AGASA
data for a  fews  eV neutrino mass in galactic or local halos. The
cross section is given by \cite{FarSal99}

\begin{multline}
\sigma_{WW}(s)=\sigma_{asym}\frac{\beta_W}{2s}\frac{1}{(s-M_Z^2)}\;\cdot \\
\cdot \left\{4 L(s) \cdot C(s)+D(s)\right\}.
\end{multline}

where $\beta_W=(1-4 M_W^2/s)^{1/2}$,
$\sigma_{asym}=\frac{\pi\alpha^2}{2\sin^4\theta_W
M_W^2}\simeq~108.5~pb$, and the functions $L(s)$, $C(s)$, $D(s)$
are defined as
\begin{displaymath}
L(s)=\frac{M_W^2}{2\beta_W s}\ln\Big(\frac{s+\beta_W s-2 M_W^2}{s-
\beta_W s-2 M_W^2}\Big)
\end{displaymath}
\begin{equation}
C(s)=s^2+s(2 M_W^2-M_Z^2)+2 M_W^2(M_Z^2+M_W^2)
\end{equation}

\begin{multline}
D(s)=\frac{1}{12 M_W^2 (s-M_Z^2)}\cdot \\
\big[ s^2(M_Z^4-60 M_W^4-4 M_Z^2 M_W^2) \\
+ 20 M_Z^2 M_W^2 s (M_Z^2+2 M_W^2) \\
-48 M_Z^2 M_W^4(M_Z^2 + M_W^2) \Big].
\end{multline}
This result should be extended with the additional new  ZZ
interaction channel considered in \cite{Far2001}:

\begin{multline}\label{4}
\sigma_{ZZ} = \frac{G^2M^2_Z}{4 \pi} y \frac{(1 + \frac{y^2}{4})}{(1 - \frac{y}{2})} \cdot~~~~~~~~ \\
\cdot \Big\{  \ln \Big[ \frac{2}{y} (1 - \frac{y}{2} + \sqrt{1 - y})
\Big] -\sqrt{1 - y} \Big\}
\end{multline}

where $y = \frac{4M^2_Z}{s}$ and $\frac{G^2M^2_Z}{4 \pi} = 35.2
\,pb$.\\

Their values are plotted in Fig.1.
 The asymptotic behaviour of
these cross section is proportional to
$\sim(\frac{M_W^2}{s})\ln{(\frac{s}{M_W^2})}$ for $s\gg M_Z^2$.\\
The nucleon arising from WW and ZZ hadronic decay could provide a
reasonable solution to the UHECR events above GZK. We'll assume
that the fraction of pions and nucleons related to the total
number of particles from the W boson decay is the almost the same
of Z boson. So W hadronic decay ($P \sim 0.68$) leads on average
to about 37 particles, where $<n_{\pi^0}> \sim 9.19$,
$<n_{\pi^{\pm}} > \sim 17$, and $<n_{p,\bar{p}, n, \bar{n}}> \sim
2.7$. In addition we have to expect by the subsequent decays of
$\pi$'s (charged and neutral), kaons and resonances ($\rho$,
$\omega$, $\eta$) produced, a flux of secondary UHE photons and
electrons. As we already pointed out, the particles resulting
from the decay are mostly prompt pions. The others are particles
whose final decay likely leads to charged and neutral pions as
well. As a consequence the electrons and photons come from prompt
pion decay.  On average it results \cite{pdg} that the energy in
the bosons decay is not uniformly distributed among the
particles. Each charged pion will give an electron (or positron)
and three neutrinos, that will have less than one per cent of the
initial W boson energy, while each $\pi^0$ decays in two photons,
each with 1 per cent of the initial W energy. In the Table~1A
above we show all the channels leading from single Z,W and Z
pairs as well as t-channel in nuclear and electro-magnetic
components.

\subsection{The process  $\nu_{i} \nu_{j} \rightarrow l_i l_j$: the t-channel}

The processes $\nu_{i} \nu_{j} \rightarrow l_i l_j$ (like
$\nu_{\mu} \nu_{\tau} \rightarrow \mu \tau $ for example)
 occur through the W boson exchange in the t-channel.
The cross-section has been derived in \cite{FarSal99}, while the
energy threshold depends on the mass of the heavier lepton
produced,\\ $E_{\nu_{th}} = 7.2 \cdot10^{19}(m_{\nu} / 0.4 \,
eV)^{-1} (m_{\tau} / m_{\tau , \mu , e})$, with the term
$(m_{\tau} / m_{\tau ,\mu , e})$ including the different
thresholds in all the possible interactions: $\nu_{\tau}
\nu_{\mu}$ (or $\nu_{\tau} \nu_e)$ , $\nu_{\mu} \nu_{e}$, and
$\nu_{e} \nu_{e}$. See Fig.2 below.

We could consider as well the reactions $\nu_{e}
\bar{\nu_{\tau_r}} \rightarrow e^- \tau^+$, $\nu_{e}
\bar{\nu_{\mu_r}} \rightarrow e^- \mu^+$ and $\nu_{e}
\bar{\nu_{e_r}} \rightarrow e^- e^+$, changing the target or the
high energy neutrino. Therefore there are  2 times more target
than for Z, WW, ZZ channels summirized in Fig.2.

 In the ultrarelativistic limit ($s \simeq
2E_{\nu} m_{\nu_r} \gg M^2_W$ where $\nu_r$ refers to relic
clustered neutrinos)  the cross-section tends to the
asymptotic value $\sigma_{\nu \bar{\nu_r}} \simeq 108.5 \,pb$.\\

\begin{multline}
 \sigma_W(s)= \sigma_{asym}\frac{A(s)}{s}
 \bigg\{1+\frac{M_W^2}{s} \cdot ~~~~~\\
 \cdot  \Bigg[2-\frac{s+B(s)}{A(s)}\ln\Bigg(\frac{B(s)+A(s)}{B(s)-
A(s)}\Bigg)\Bigg]\bigg\}
\end{multline}
where $\sqrt{s}$ is the center of mass energy, the functions
A(s), B(s) are defined as
\begin{multline}
A(s)= \sqrt{[s-(m_\tau+m_\mu)^2] [s-(m_\tau-m_\mu)^2]} \\
B(s)=s+2M_W^2-m_\tau^2-m_\mu^2 ~~~~~~~~
\end{multline}
and
\begin{equation}
\sigma_{asym}=\frac{\pi\alpha^2}{2\sin^4\theta_W
M_W^2}\simeq~108.5~pb
\end{equation}
where $\alpha$ is the fine structure constant and $\theta_W$ the
Weinberg angle; $\sigma_{asym}$ is the asymptotic behaviour of
the cross section in the ultra-relativistic limit

\begin{equation}
s\simeq 2 E_\nu m_\nu= 2\cdot 10^{23} \frac {E_\nu}{10^{22}~eV}  \frac {m_\nu} {10~eV}~eV^2 \gg M_W^2~.
\end{equation}
\begin{figure}[h]
\begin{center}
 \includegraphics[width=0.45\textwidth] {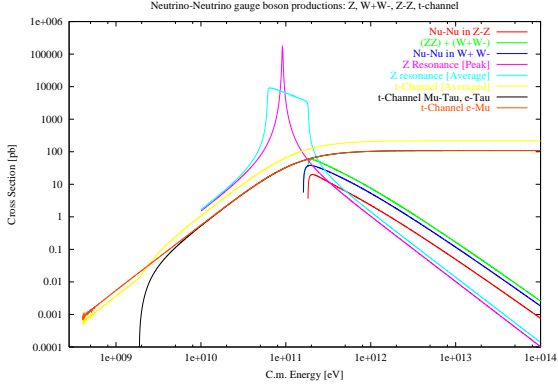}
\end{center}
  \caption{The  $\nu \bar{\nu} \rightarrow  Z,W^+ W^-,ZZ,T$-channel,  cross sections as a function of the center of mass
energy in $\nu \nu$.
   These cross-sections are estimated also in average (Z) as well for each possible
   t-channel lepton pairs. The averaged t-channel averaged the multiplicity of flavours
    pairs ${\nu}_{i}$, $ \bar{\nu}_{j}$ respect to neutrino
    pair annihilations into Z neutral boson. The Z-WW-ZZ Showering
    has to be relativistically boosted to show their behaviour at laboratory system.} \label{fig:boxed_graphic 1}
\end{figure}
This interactions, as noted in Table~1A are leading to
electro-magnetic showers and are not offering any nuclear
secondary.

\section{The Boosted Z-UHECR  spectra }

Let us examine the destiny of UHE primary particles (nucleons,
electrons and photons) ($E_e \lesssim 10^{21}\,eV$) produced
after hadronic or leptonic W decay. As we already noticed in the
introduction, we'll assume that the nucleons, electrons and
photons spectra (coming from W or Z decay) after $\nu \nu$
scattering in the halo, follow a power law that in the center of
mass system is $\frac{dN^*}{dE^* dt^*} \simeq E^{* - \alpha}$
where $\alpha \sim 1.5$. This assumption is based on detailed
Monte Carlo simulation of a heavy fourth generation neutrino
annihilations \cite{Konoplich} \cite{Konoplich2} \cite{Grossi}
and with the model of quark - hadron fragmentation spectrum
suggested by Hill \cite{Hill}.

In order to determine the shape of the particle spectrum in the
laboratory frame, we have to introduce the Lorentz relativistic
transformations from the center of mass system to the laboratory
system.
 The number of particles is clearly a relativistic invariant $dN_{lab} = dN^*$,
while the relation between the two time intervals is $dt_{lab} =
\gamma dt^*$, the energy changes like $ \epsilon_{lab} = \gamma
\epsilon^* (1 + \beta \cos \theta^*) = \epsilon^* \gamma^{-1}(1 -
\beta \cos \theta)^{-1}$, and finally the solid angle in the
laboratory frame of reference becomes $d\Omega_{lab} =\gamma^{2}
d\Omega^*  (1 - \beta \cos \theta )^2$. Substituting these
relations one obtains


\begin{multline}
\left(\frac{dN}{d\epsilon dt d\Omega} \right)_{lab} =
\frac{dN_{*}}{d\epsilon_{*} dt_{*} d\Omega_{*}} \gamma^{-2}
 (1 - \beta \cos \theta)^{-1} \\
=\frac{\epsilon^{-\alpha}_{*} \; \gamma^{-2}} {4 \pi} \cdot (1 - \beta \cos
 \theta)^{-1} \\
= \frac{\epsilon^{-\alpha} \; \gamma^{-\alpha-2}} {4 \pi} (1 -
\beta \cos \theta)^{-\alpha-1}
\end{multline}

and integrating on $\theta$ (omitting the lab notation) one loses
the spectrum dependence on the angle.


The consequent fluence derived by the solid angle integral is:

\begin{multline}
\frac{dN}{d\epsilon dt} \epsilon^{2}= \\
 \frac{\epsilon^{-\alpha+2} \; \gamma^{\alpha-2}} {2 \beta \alpha}
 [(1 + \beta)^{\alpha} - (1 - \beta)^{\alpha}] \simeq \\
 \simeq \frac{2^{\alpha-1}\epsilon^{-\alpha+2} \; \gamma^{\alpha-2}} {\alpha}
\end{multline}

There are two extreme case to be considered: the case where the
interaction occurs at Z peak resonance and therefore the center of
mass Lorents factor $\gamma$ is frozen at a given value (eq.1)
and the case (WW,ZZ pair channel) where all energies are
allowable  and $\gamma$ is proportional to $\epsilon^{1/2}$.
 Here we focus only on Z peak resonance. The consequent fluence spectra
 $\frac{dN}{d\epsilon dt}\epsilon^{2}$, as above, is proportional to $\epsilon^{-\alpha +2}$. Because $\alpha$ is
nearly $1.5$ all the consequent secondary particles will also show
a spectra proportional to $\epsilon^{1/2}$ following a normalized
energies shown in Tab.2, as shown in Fig.(2-6). In the latter
case (WW,ZZ pair channel), the relativistic boost reflects on the
spectrum of the secondary particles, and the spectra power law
becomes $\propto \epsilon^{\alpha/2 +1}=\epsilon^{0.25}$. These
channels will be studied in details elsewhere. In Fig.~1 we show
the spectrum of protons, photons and electrons coming from Z
hadronic and leptonic decay assuming a nominal primary CR energy
flux $\sim 20~eV s^{-1} sr^{-1} cm^{-2}$, due to the total $\nu
\bar{\nu}$ scattering at GZK energies as shown in figures 2-6.
Let us remind that we assume an interaction probability of $\sim
1 \%$ and a corresponding UHE incoming neutrino energy $\sim
2000~eV s^{-1} sr^{-1} cm^{-2}$ near but below present $UHE$
neutrino flux bound from AMANDA and Baikal as well as Goldstone
data.

\vspace{0.3cm}
\begin{table}[h]
\begin{tabular}{ccc}
 \cline{1-3}
 \vspace{-0.6cm} &  & \\ \cline{1-3}
\multicolumn{3}{c}{\vspace{-0.2cm}SECONDARIES ENERGY DISTRIBUTIONS} \\
\multicolumn{3}{c}{
In Z Decay ($m_{\nu}=0.4 \; eV$)} \\
\cline{1-3}
\vspace{-0.6cm} &  & \\ \cline{1-3}
  $~~~~Channel~~~~$ & $E (eV)$ & $\frac{dN}{dE}E^2$ (eV) \\ \cline{1-3}
  $p$ &  $2.2 \cdot 10^{20}$ & 1.2 \\ \cline{1-3}
  $\gamma$ & $9.5 \cdot 10^{19}$ & 4.25 \\ \cline{1-3}
  $e_{\pi}$ & $5 \cdot 10^{19}$ & 2.3 \\ \cline{1-3}
   $e_{prompt}$ & $5 \cdot 10^{21}$ & 1.32 \\ \cline{1-3}
    $e_{\mu}$ & $1.66 \cdot 10^{21}$ & 0.45 \\ \cline{1-3}
     $e_{\tau}$ & $1.2 \cdot 10^{21}$ & 0.6 \\ \cline{1-3}
\vspace{-0.6cm} &  & \\ \cline{1-3}
\end{tabular}

\caption{B. Energy peak and Energy Fluence for different decay
channels as described in the text.We assumed that in the centre
of mass frame, the energy of the proton and of the pion are
respectively described in Fig.1 Table 1A}
\end{table}

\section{The UHECRs from Relic $\nu$ Masses}

The role of each relic neutrino mass is summirized from the
convolutions of the UHE neutrino spectra with the relic neutrino
mass, its density as well as the cross-sections described above.
The case of Z-resonance event with a single neutrino mass has a
narrow fine tuned energy mass windows (0.4 eV-1.2 eV) described
respectively in Figures 3-4.

\begin{figure}[h]
\begin{center}
 \includegraphics[width=0.45\textwidth] {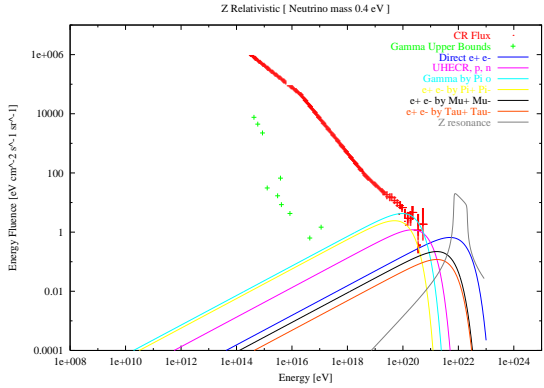}
\end{center}
  \caption{Energy Fluence derived by $\nu \bar{\nu} \rightarrow Z$ and its showering into
  different channels: direct electron pairs UHECR nucleons $n$ $p$ and anti-nucleons, $\gamma$ by $\pi^0$ decay,
  electron pair by $\pi^+ \pi^-$ decay, electron pairs by direct muon and tau decays as labeled in figure.
  The relic neutrino mass has been assumed to be fine tuned to explain GZK UHECR tail:
  $m_{\nu}=0.4 eV$. The Z resonance ghost (the shadows of Z Showering resonance \cite{Far2001} curve),
  derived from Z cross-section
  in Fig.1, shows the averaged $Z$ resonant cross-section peaked
  at $E_{\nu}=10^{22} eV$. Each channel shower has been normalized following table 1B.}
\label{fig:boxed_graphic 2}
\end{figure}

\begin{figure}[h]
\begin{center}
 \includegraphics[width=0.45\textwidth] {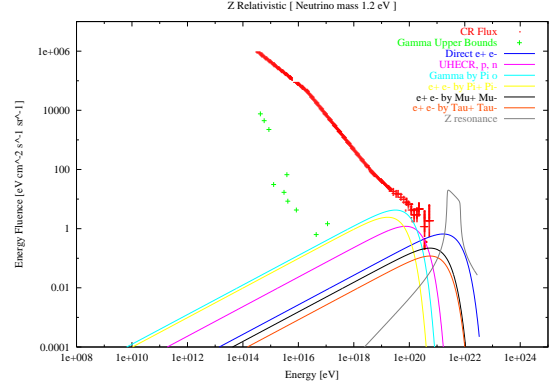}
\end{center}
  \caption{Energy Fluence derived by $\nu \bar{\nu} \rightarrow Z$ and its showering into
  different channels as in previous Figure 2: direct electron pairs UHECR nucleons $n$ $p$, $\gamma$ by $\pi^0$ decay,
  electron pair by $\pi^+ \pi^-$ decay, electron pairs by direct muon and tau decays as labeled in figure.
  In the present case the relic neutrino mass has been assumed to be fine tuned to explain GZK UHECR tail:
  $m_{\nu}=1.2 eV$ with the same UHE incoming neutrino fluence of previous figure. The Z resonance curve shows the averaged $Z$ resonant cross-section peaked
  at $E_{\nu}=3.33\cdot10^{21} eV$.Each channel shower has been normalized in analogy to table 1B.}
  \label{fig:boxed_graphic 3}
\end{figure}
We remind again that a heavier neutrino mass $(\geq 2 eVs)$ imply
the rise of WW-ZZ channels and a pile up of Z resonance
cross-section at lower UHECR spectra. This feature maybe already
responsible for the tiny bump in observed events around
$5\cdot10^{19}eV$. The lighter neutrino mass  possibilities (near
0.1 eV) are comparable with present Super-Kamiokande atmospheric
neutrino mass and are leading to the exciting scenario where more
non degenerated Z-resonances occur \cite{Far2001}. These scenario
are summarized in Fig. 5 (for nominal example $m_{\nu_{\tau}}$ =
0.1 eV; $m_{\nu_{\mu}}$ = 0.05 eV). The twin neutrino mass inject
a corresponding twin bump at highest energy. Another limiting case
of interest takes place when the light neutrino masses are
extreme, nearly at atmospheric (SK,K2K) and solar (SNO) neutrino
masses. This case is described in two different versions in Fig.6
(assuming comparable neutrino densities) and Fig.7 (keeping care
of the lightest neutrino density diluitions). The relic neutrino
masses are assumed $m_{\nu_{\tau}}$ = 0.05 eV; $m_{\nu_{\mu}}$ =
0.001 eV). A more complex scenario is also possible when it takes
place both a narrow twin bump (Fig5) $and$ a wider twin bump (Fig
6-7) because of a small neutrino tau-muon mass splitting
overlapping with a wider one due to lightest neutrino electron
mass.

\begin{figure}[h]
\begin{center}
\includegraphics[width=0.45\textwidth] {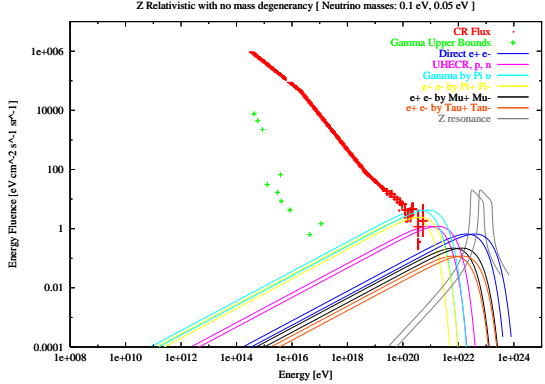}
\end{center}
  \caption{Energy Fluence derived by $\nu \bar{\nu} \rightarrow Z$ and its showering into
  different channels: direct electron pairs UHECR nucleons $n$ $p$, $\gamma$ by $\pi^0$ decay,
  electron pair by $\pi^+ \pi^-$ decay, electron pairs by direct muon and tau decays as labeled in figure.
  In the present case the relic neutrino masses have been assumed with no degenerancy.
  Their values have been fine tuned to explain GZK UHECR tail:
   $m_{\nu_1}=0.1 eV$ and $m_{\nu_2}=0.05 eV$. No relic neutrino
   density difference has been assumed.
   The incoming UHE neutrino fluence has been increased
   by a factor 2 respect previous Fig.3-4. The Z resonance curve shows the averaged $Z$ resonant cross-section peaked
  at $E_{\nu_1}=4\cdot10^{22} eV$ and $E_{\nu_2}=8\cdot10^{22} eV$. Each channel shower has been normalized in analogy to table 1B.}
\label{fig:boxed_graphic 4} 
\end{figure}

\begin{figure}[h]
\begin{center}
 \includegraphics[width=0.45\textwidth] {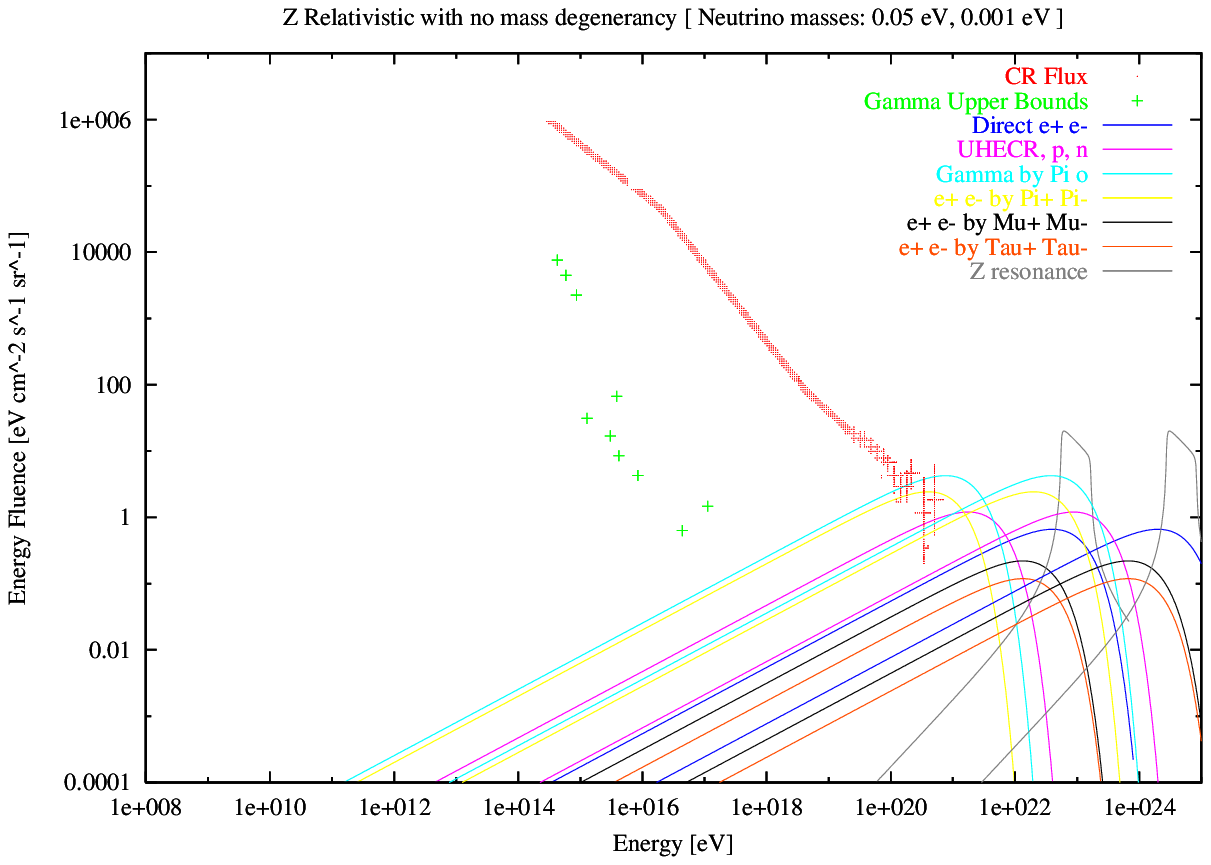}
\end{center}
\caption{Energy Fluence derived by $\nu \bar{\nu} \rightarrow Z$
and its showering into
  different channels  as above.
  In the present extreme case the relic neutrino masses have been assumed with wide mass differences
  just compatible both with Super-Kamiokande and relic $2 K^{o}$ Temperature.
  The their values have been fine tuned to explain observed GZK- UHECR tail:
   $m_{\nu_1}=0.05eV$ and $m_{\nu_2}=0.001 eV$. No relic neutrino
   density difference between the two masses  has been assumed,
   contrary to bound in eq.3. The incoming UHE neutrino fluence has been increased
   by a factor 2 respect previous Fig.2-3. The "Z resonance" curve
    shows the averaged $Z$ resonant cross-section peaked
  at $E_{\nu_1}=8\cdot10^{22} eV$ and $E_{\nu_2}=4\cdot10^{24} eV$, just
  near Grand Unification energies. Each channel shower has been normalized in analogy to table 1B.}
\label{fig:boxed_graphic 5}
\end{figure}

\begin{figure}[h]
\begin{center}
 \includegraphics[width=0.45\textwidth] {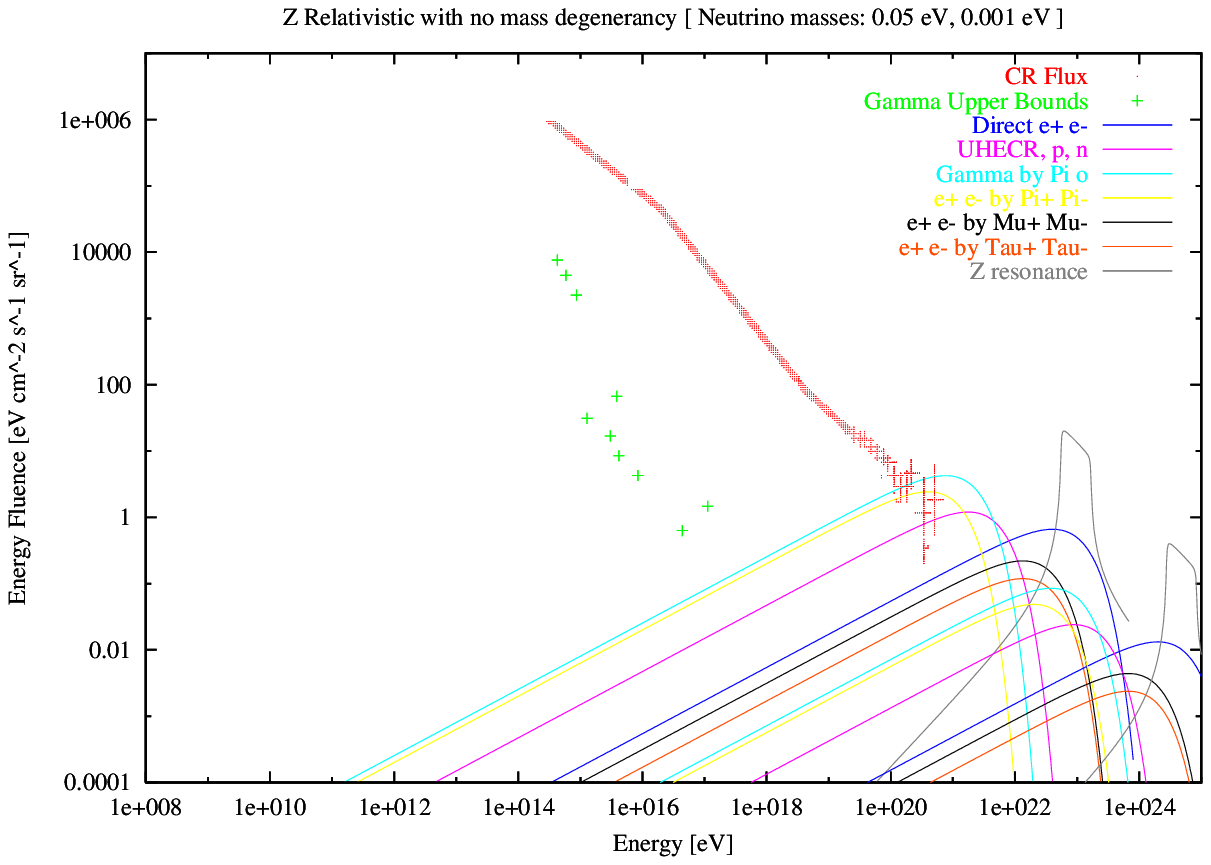}
\end{center}
\caption{Energy Fluence derived by $\nu \bar{\nu} \rightarrow Z$
and its showering into  different channels  as above.
  In the present extreme case the relic neutrino masses have been assumed with wide mass differences
  just compatible both with Super-Kamiokande and relic $2 K^{o}$ Temperature.
  The their values have been fine tuned to explain observed GZK- UHECR tail:
   $m_{\nu_1}=0.05eV$ and $m_{\nu_2}=0.001 eV$. A neutrino
   density difference between the two masses  has been
   assumed,considering the lightest $m_{\nu_2}=0.001 eV$ neutrino
   at relativistic regime, consistent to bound in eq.3.
   The incoming UHE neutrino fluence has been assumed growing
   linearly \cite{Yoshida}  with energy. Its value is increased
   by a factor 2 and 20  at   $E_{\nu_1}=8\cdot10^{22} eV$ and $E_{\nu_2}=4\cdot10^{24} eV$
   respect the previous ones Fig.2-3. The "Z resonance" curve
    shows its averaged $Z$ resonant "ghost" cross-section peaked
  at $E_{\nu_1}=2\cdot10^{23} eV$ and $E_{\nu_2}=4\cdot10^{24} eV$, just
  near Grand Unification energies. Each channel shower has been normalized in analogy to table 1B.}
\label{fig:boxed_graphic 6}
\end{figure}

\section{UHECRs  Anisotropy and Clustering}

The  neutrino mass play a role in defining its Hot Dark Halo size
and the consequent enhancement of UHECR arrival directions due to
our peculiar position in the HDM halo. Indeed for a heavy $\geq 2
eV$ mass case HDM neutrino halo are mainly galactic and/or local,
reflecting an isotropic or a diffused amplification toward nearby
$M31$ HDM halo. In the lighter case the HDM should include the
Local Cluster up to Virgo. To each size corresponds also a
different role of UHECR arrival time. The larger the HDM size the
longer the UHECR random-walk travel time (in extra-galactic
random magnetic fields) and the wider the arrival rate lag
between doublets or triplets. The smaller is the neutrino halo
the earlier the UHE neutron secondaries by Z shower will play a
role: indeed at $E_{n}= 10^{20}eV$ UHE neutron are flying a Mpc
and their directional arrival (or their late decayed proton
arrival) are more on-line toward the source. This may explain the
high self collimation and auto-correlation of UHECR discovered
very recently \cite{Tinyakov}. The UHE neutrons Z-Showering fits
with the harder spectra observed in clustered events in AGASA
\cite{Takeda}. The same UHECR alignment may explain the quite
short (2-3 years)\cite{Takeda2} lapse of time observed in AGASA
doublets. Indeed the most conservative scenario where UHECR are
just primary proton from nearby sources at GZK distances (tens of
Mpcs) are no longer acceptable either because the absence of such
nearby sources and because of the observed stringent UHECR
clustering ($2^o - 2.5^o$) \cite{Takeda} in arrival direction, as
well as because of the short ($\sim3$ years) characteristic time
lag between clustered events. Finally the same growth with energy
of UHECR neutron (and anti-neutron) life-lengths (while being
marginal or meaning-less in tens Mpcs GZK flight distances) may
naturally explain, within a $\sim Mpc $ Z Showering Neutrino
Halo, the arising harder spectra revealed in doublets-triplet
spectra \cite{Aoki}.

\section{The Tinyakov-Glushkov  Paradox }

The same role of UHE neutron secondaries from Z showering in HDM
halo may also solve an emerging puzzle: the  correlations of
arrival directions of UHECRs found recently \cite{Glushkov} in
Yakutsk data at energy $E= 8\cdot 10^{18} eV$ toward the Super
Galactic Plane are to be compared with the compelling evidence of
UHECRs events ($E= 3\cdot 10^{19} eV$ above GZK) clustering toward
well defined BL Lacs at cosmic distances (redshift $z> 0.1-0.2$)
\cite{Tinyakov,Tinyakov2}. Where is the real UHECR sources
location? At Super-galactic disk (50 Mpcs wide, within GZK range)
or at cosmic ($\geq 300Mpcs$) edges? It should be noted that even
for the Super Galactic hypothesis \cite{Glushkov} the common
proton are unable to justify the high collimation of the UHECR
events. Of course both results (or just one of them) maybe a
statistical fluctuation. But both studies seem statistically
significant (4.6-5 sigma) and they seem in obvious disagreement.
There may be still open the  possibility of $two$ new categories
of UHECR sources both of them located at different distances
above GZK ones (the harder the most distant BL Lac sources). But
it seem quite unnatural  the UHECR propagation by direct
nucleons   where the most distant are the harder. However our
Z-Showering scenario offer different solutions: (1) The Relic
Neutrino Masses define different Hierarchical Dark Halos and
privileged arrival direction correlated to Hot Relic Neutrino
Halos. The real sources are at (isotropic) cosmic edges
\cite{Tinyakov}, \cite{Tinyakov2}, but their crossing along a
longer anisotropic relic neutrino cloud enhance the interaction
probability in the Super Galactic Plane. (2) The nearest SG
sources are weaker while the collimated BL Lacs are harder: both
sources need a Neutrino Halo to induce the Z-Showering UHECRs.
More data will clarify better the real scenario.

\section{ The TeV Tails from UHE electrons }

 As it is shown in Table 1A-B and Figures above, the electron
(positron) energies by $\pi^{\pm}$ decays is around $E_e \sim 2
\cdot 10^{19} \, eV$ for an initial $E_Z \sim 10^{22} \, eV $ (
and $E_{\nu} \sim 10^{22} \, eV $). Such electron pairs while not
radiating efficently in extra-galactic magnetic fields will be
interacting with the galactic magnetic field ($B_G \simeq 10^{-6}
\,G $)  leading to direct TeV photons:
\begin{displaymath}
  E_{\gamma}^{sync} \sim \gamma^2 \left( \frac{eB}{2\pi m_e } \right)
   \sim
\end{displaymath}
\begin{equation}\label{4b}
  \sim 27.2 \left( \frac{E_e}{2 \cdot10^{19}
  \,eV} \right)^2 \left( \frac{m_{\nu}}{0.4 \, eV} \right)^{-2} \left( \frac{B}{\mu G} \right)\,TeV.
\end{equation}
The same UHE electrons will radiate less efficiently with extra-
galactic magnetic field ($B_G \simeq 10^{-9} \,G $)  leading also
to direct peak $27.2$ GeV  photons.
   The spectrum of these photons is characterized by a power of law $dN
/dE dT \sim E^{-(\alpha + 1)/2} \sim E^{-1.25}$ where $\alpha$ is
the power law of the electron spectrum, and it is showed in
Figures above. As regards the prompt electrons at higher energy
($E_e \simeq 10^{21}\, eV$), in particular in the t-channels,
their interactions with the extra-galctic field first and
galactic magnetic fields later is source of another kind of
synchrotron emission around tens of PeV energies
$E^{sync}_{\gamma}$:

\begin{equation}\label{2}
 \sim
  6.8 \cdot 10^{13} \left( \frac{E_e}{10^{21}\,eV} \right)^2
  \left( \frac{m_{\nu}}{0.4 \, eV} \right)^{-2} \left( \frac{B}{nG}
  \right) \, eV
\end{equation}
\begin{equation}\label{2b}
 \sim
  6.8 \cdot 10^{16} \left( \frac{E_e}{10^{21}\,eV} \right)^2
  \left( \frac{m_{\nu}}{0.4 \, eV} \right)^{-2} \left( \frac{B}{\mu G}
  \right) \, eV
\end{equation}
The corresponding energy loss length instead is \cite{Ka}
\begin{equation}\label{3}
\left( \frac{1}{E} \frac{dE}{dt} \right)^{-1} = 3.8 \times \left(
\frac{E}{10^{21}} \right)^{-1} \left( \frac{B}{10^{-9} G}
\right)^{-2} \, kpc.
\end{equation}
For the first case the interaction lenght is few Kpcs while in
the second one in few days light flight. Again one has the same
power law characteristic of a synchrotron spectrum with index
$E^{-(\alpha + 1 / 2)} \sim E^{-1.25}$.
 Gammas at $10^{16} \div 10^{17}$ eV scatters onto
low-energy photons from isotropic cosmic background ($\gamma + BBR
\rightarrow e^+ e^-$) converting their energy in electron pair.
 The expression of the pair production cross-section is:
\begin{equation} \sigma (s) = \frac{1}{2} \pi r_0^2 (1 - v^2) [
(3 - v^4) \ln \frac{1 + v}{1 - v} - 2 v (2 - v^2) ]
\end{equation}
where $v = (1 - 4m_e^2 / s)^{1/2}$,  $s = 2 E_{\gamma} \epsilon (1
- \cos \theta)$ is the square energy in the center of mass frame,
$\epsilon$ is the target photon energy, $r_0 $  is the classic
electron radius, with a peak cross section value at
\[ \frac{4}{137}\times \frac{3}{8\pi} \sigma_T \ln 183 = 1.2 \times
10^{-26} \,cm^2 \] Because the corresponding attenuation length
due to the interactions with the microwave background is around
ten kpc, the extension of the halo plays a fundamental role in
order to make this mechanism efficient or not. As is shown in
Fig.3-4, the contribution to tens of PeV gamma signals by Z (or
W) hadronic decay, could be compatible with actual experimental
limits fixed by CASA-MIA detector on such a range of energies.
Considering a halo extension $l_{halo} \gtrsim 100 kpc$, the
secondary electron pair creation becomes efficient, leading to a
suppression of the tens of PeV signal. So electrons at $E_e \sim
3.5 \cdot 10^{16} \,eV$ loose again energy through additional
synchrotron radiation\cite{Ka}, with maximum $E_{\gamma}^{sync}$
around
\begin{equation}\label{3b}
  \sim 79 \left( \frac{E_e}{10^{21}
  \,eV} \right)^4 \left( \frac{m_{\nu}}{0.4 \, eV} \right)^{-4}
   \left( \frac{B}{\mu G} \right)^3 \, MeV.
\end{equation}
Anyway this signal is not able to pollute sensibly the MeV-GeV;
the relevant signal pile up at TeVs.

Gamma rays with energies up to 20 TeV have been observed by
terrestrial detector only by nearby sources like Mrk 501 (z =
0.033) or very recently by MrK 421. This is puzzling because the
extra-galactic TeV spectrum should be, in principle, significantly
suppressed by the $\gamma$-rays interactions with the
extra-galactic Infrared background, leading to electron pair
production and TeVs cut-off. The recent calibration and
determination of the infrared background by DIRBE and FIRAS on
COBE have inferred severe constrains on TeV propagation. Indeed,
as noticed by Kifune \cite{Kifune}, and Protheroe and
Meyer\cite{Meyer} we may face a severe infrared background - TeV
gamma ray crisis. This crisis imply a distance cut-off,
incidentally, comparable to the GZK one. Let us remind also an
additional evidence for IR-TeV cut-off is related to the possible
discover of tens of TeV counterparts of BATSE GRB970417, observed
by Milagrito\cite{Milagrito}, being most GRBs very possibly at
cosmic edges, at distances well above the IR-TeV cut-off ones. In
this scenario it is also important to remind the possibilities
that the Fly's Eye event has been correlated to TeV pile up
events in HEGRA \cite{Horns}. The very recent report (privite
communication 2001) of the absence of the signal few years  later
at HEGRA may be still consistent with a bounded Z-Showering
volume and a limited UHE TeV tail activity.
 To solve the IR-TeV cut-off one may alternatively invoke unbelievable extreme hard intrinsic
spectra or exotic explanation as gamma ray superposition of
photons or sacrilegious  Lorentz invariance violation
\cite{Camelia}.
\begin{figure}[h]
\begin{center}
 \includegraphics[width=0.45\textwidth] {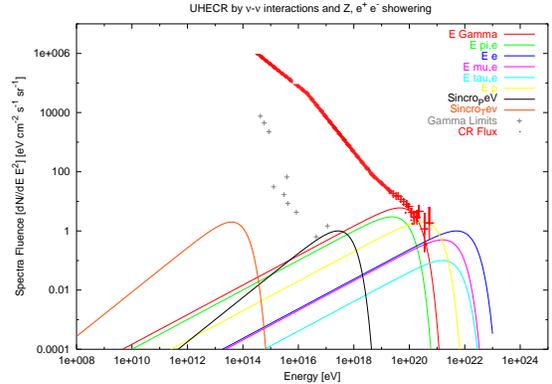}
\end{center}
\caption{Energy fluence by Z showering as in fig.3 and the
consequent $e^+ e^-$ synchrotron radiation  by eq.16-18}
\label{fig:boxed_graphic 7}
\end{figure}

\begin{figure}[h]
\begin{center}
 \includegraphics[width=0.45\textwidth] {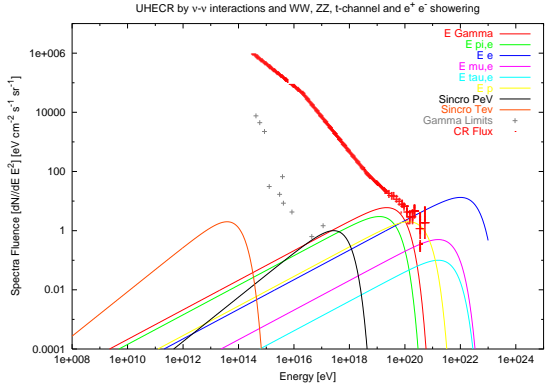}
\end{center}
\caption{Energy fluence by WW, ZZ, t-channel showering as in fig.3
and the consequent $e^+ e^-$ synchrotron radiation  by eq.16-18.
The lower energy Z showering is not included to make spectra more
understandable.} \label{fig:boxed_graphic 8}
\end{figure}

\section{Conclusion}

UHECR above GZK may be naturally born by UHE $\nu$ scattering on
relic ones.  The target cosmic $\nu$ may be light and dense as
the needed ones in HDM model (few eVs). Then their $W^+ W^-,ZZ$
pair productions channel (not just the  Z resonant peak) would
solve the GZK puzzle. At a much lighter, but fine tuned case
$m_{\nu}\sim 0.4 eV$, $m_{\nu}\sim 1.5 eV$ assuming $E_{\nu}\sim
10^{22} eV$, one is able to solve at once the known UHECR data at
GZK edge by the dominant Z peak; in this peculiar scenario one may
foresee  (fig.2-3) a rapid   decrease (an order of magnitude in
energy fluence) above $3\cdot10^{20}eV$ in future data
   and a further recover (due to WW,ZZ channels) at higher energies.
   The characteristic UHECR fluxes will reflect the averaged
   neutrino-neutrino interactions shown in Fig.2-7.
      Their imprint could confirm the neutrino masses value and relic
   density. At a more extreme lighter neutrino mass, occurring for
$m_{\nu}\sim m_{\nu_SK}\sim 0.05 eV$, the minimal
$m_{\nu_{\tau}},m_{\nu_{\mu}}$ small mass differences might be
reflected, in a spectacular way, into UHECR modulation quite above
the GZK edges.  The "twin" lightest masses (Fig.5-6-7) call for
either gravitational $\nu$ clustering above the expected one
 or the presence of relativistic diffused background.
Possible neutrino gray body spectra, out of thermal equilibrium,
at higher energies may also arise from non standard early
Universe. The UHECR acceleration is not yet solved, but their
propagation from far cosmic volumes is finally allowed.
 The role of UHE neutrons in Z-showering, their directional flight leading to clustering in
  self collimated  data is possibly emerging by harder spectra.
 Peculiar secondaries of TeVs tails may be precursor and
 afterglows signal correlated to past or future UHECRs
pointing toward the same far sources. The IR-TeV solution may be
just be a necessary corollary of the Z-Showering GZK solution
\cite{Fargion3}. The  time and space directional may be a new
fundamental test of present  Z-Showering model. The discover of
UHE neutrino at GZK energies might be testify on ground by UHE
$\tau$ air-shower, born by direct $10^{19}$eV UHE $\nu$ crossing
small Earth crust depth, flashing from the horizontal edges to
mountain,balloon and satellite detectors \cite{Fargion2}.
 The new generation  UHECR data within next decade,  may also offer the
 probe of lightest elementary particle masses, their
relic densities, their spatial map distribution and energies and
the most ancient and evasive shadows of earliest $\nu$ cosmic
relic backgrounds.

\end{document}